\documentclass{ws-mpla}
\usepackage{url}
\usepackage{subfigure}
\usepackage{graphicx,color,dcolumn}
\usepackage{url}
\RequirePackage{color}

\begin{document}

\markboth{Authors' Names}
{Instructions for Typing Manuscripts (Paper's Title)}

\catchline{}{}{}{}{}

\title{MODIFIED FERMI ENERGY OF ELECTRONS IN A SUPERHIGH MAGNETIC FIELD }
\author{\footnotesize CUI, ZHU}
\address{1. Xinjiang Astronomical Observatory, CAS, Urumqi, Xinjiang, 830011, China. \\
2. Key Laboratory of Radio Astronomy, CAS, Nanjing, Jiangshu, 210008, China \\
3.University of Chinese Academy of Sciences, Beijing, 100049, China }
\author{\footnotesize ZHI FU, GAO}
\address{1. Xinjiang Astronomical Observatory, CAS, Urumqi, Xinjiang, 830011, China. zhifugao@xao.ac.cn\\
2. Key Laboratory of Radio Astronomy, CAS, Nanjing, Jiangshu, 210008, China }
\author{XIANG DONG, LI}
\address{1. Shchool of Astronomy and Space Science, Nanjing University, Nanjing, Jiangsu, 210093, China \\
2. Key Laboratory of Modern Astronomy and Astrophysics, Ministry of Education, Nanjing University, Jiangsu, China}
\author{NA, WANG}
\address{Xinjiang Astronomical Observatory, CAS, Urumqi Xinjiang, 830011, China}
\author{JIAN PING, YUAN}
\address{Xinjiang Astronomical Observatory, CAS, 150, Science
1-Street, Urumqi, Xinjiang, China}
\author{ QIU HE, PENG}
\address{Shchool of Astronomy and Space Science, Nanjing University, Nanjing, Jiangsu, 210093, China}

\maketitle

\pub{Received (Day Month Year)}{Revised (Day Month Year)}

\begin{abstract}
In this paper, we investigate the electron Landau-level stability and its influence
on the electron Fermi energy, $E_{\rm F}(e)$, in the circumstance of magnetars, which are
powered by magnetic field energy. In a magnetar, the Landau levels of degenerate and
relativistic electrons are strongly quantized. A new quantity $g_{n}$, the electron
Landau-level stability coefficient is introduced. According to the requirement that
$g_{n}$ decreases with increasing the magnetic field intensity $B$, the magnetic-field
index $\beta$ in the expression of $E_{\rm F}(e)$ must be positive. By introducing the
Dirac$-\delta$ function, we deduce a general formulae for the Fermi energy of
degenerate and relativistic electrons, and obtain a particular solution to $E_{\rm F}(e)$
in a superhigh magnetic field (SMF). This solution has a low magnetic-field index of
$\beta=1/6$, compared with the previous one, and works when $\rho\geq 10^{7}$~g cm$^{-3}$
and $B_{\rm cr}\ll B\leq 10^{17}$~Gauss. By modifying the phase space of relativistic
electrons, a SMF can enhance the electron number density $n_e$,
and decrease the maximum of electron Landau level number, which results in a
redistribution of electrons. According to Pauli exclusion principle, the degenerate
electrons will fill quantum states from the lowest Landau level to the
highest Landau level. As $B$ increases, more and more electrons will occupy higher
Landau levels, though $g_{n}$ decreases with the Landau level number $n$. The
enhanced $n_{e}$ in a SMF means an increase in the electron
Fermi energy and an increase in the electron degeneracy pressure. The results are
expected to facilitate the study of the weak-interaction processes inside neutron
stars and the magnetic-thermal evolution mechanism for megnetars.
\end{abstract}
\ccode{PACS: 97.60.Jd; 21.65.-f;71.18.+y}
\section{Introduction}	
It is universally recognized that pulsars are highly magnetized neutron stars (NSs),
with surface dipole magnetic field being about $10^{10}-10^{12}$ Gauss. Megnetars are a kind of
pulsars powered by their magnetic energy rather than their rotational energy, and
their surface dipole magnetic fields are generally 2$-$3 orders of magnitude higher
than those of radio pulsars ($B^{*}= B/B_{\rm cr}\gg 1$, $B_{\rm cr}= 4.414 \times 10^{13}$ Gauss
is the quantum critical field of electrons), and their internal magnetic
fields might be even higher (e.g., see
Ref.\refcite{Thompson96}). Megnetars are categorized into
two populations historically: Soft Gamma$-$ray Repeaters (SGRs) and Anomalous X$-$ray
Pulsars (AXPs). The SGR flares were explained as resulting from violent magnetic
reconnections and crustal quakes, and the quiescent X-ray emission of AXPs
(with X-ray luminosities much larger than their spin-down luminosities) was
attributed to the decay of superhigh magnetic fields (e.g., see
Ref.~\refcite{Colpi00},~\refcite{Woods04}) under which the
Landau levels of electrons are strongly quantized.

For the completely degenerate and relativistic electrons in $\beta-$equilibrium,
the distribution function $f(E_{e})$ obeys Fermi-Dirac statistics (see Ref.~\refcite{Gao13}).
When the temperature $T\rightarrow 0$, the electron chemical potential $\mu_{e}$ is also called ``the electron Fermi
energy", $E_{\rm F}(e)$, which has the simple form of $E_{\rm F}^2{}(e)= p^{2}_{\rm F}(e)c^{2}+ m^{2}_{e}c^{4}$,
with $p_{\rm F}(e)$ being the electron Fermi momentum. As an extremely
important and indispensable physical parameter in the equation of state (EoS) of a NS,
the Fermi energy of electrons directly exerts impact on the weak-interactions
processes, including modified Urca reactions, $\beta-$decay, electron capture, as well
as the absorption of neutrinos and anti-neutrinos, etc (see Ref.~\refcite{Lam14},~\refcite{Martinez-Pinedo14}). They
will in turn influence the intrinsic EoS, internal structure, thermal evolution, and
even the overall properties of the star (see Ref.~\refcite{Li16}). Therefore, it is of great significance
to study $E_{\rm F}(e)$ in the circumstance of a NS.

Since $E_{\rm F}(e)$ increases with the increasing in the depth of a
NS (see Ref.~\refcite{Shapiro83}), it is necessary to briefly review the structure of
the star. The structure of a NS roughly includes an atmosphere and four major internal
regions: the outer crust, inner crust, outer core, and inner core.
The outer shell comprises crystal lattices and electrons, which are distributed from the surface
of the star to the region where the neutron-drop density (see Ref.~\refcite{BPS71}) $\rho_{d}$ is reached.  At the
point $\rho=\rho_{d}$, the neutrons begin to overflow from the nuclei, forming a free neutron gas, and the
value of $E_{\rm F}(e)$ is about 25 MeV (see Ref.~\refcite{BPS71},~\refcite{Ruster06}).
The inner shell is mainly composed of degenerate and
relativistic electrons, non-relativistic nucleons and
over-saturated neutrons, distributed from the region of neuron-drop density to
the shell-core boundary ($\rho\sim 0.5\rho_{0}$, $\rho_{0}=2.8\times 10^{14}$~g~cm$^{-3}$ is the
standard nuclear density). Nuclei fully disappear at this density, and $E_{\rm F}(e)$
is about 35 MeV. The outer core is composed of neutrons and a small amount of electrons and
protons, with the density range being $0.5\rho_{0}\sim 2.5\rho_{0}$. The Fermi energy of
electrons in this region is estimated as $E_{\rm F}(e)=60\times(\rho/\rho_{0})
^{2/3}$~MeV (see Ref.~\refcite{Shapiro83}). For the region with higher density, the electron Fermi
energy could exceed the muon rest-mass energy $m_{\mu}c^2=105.7$ MeV, and a small amount of muons($\mu$) are detected.
The inner core is about several kilometers in radius, and has a central density as high as $\sim$
$10^{15}$~g~cm$^{-3}$. When $\rho>\rho_{tr}$, some nucleons will transform to ¡®exotic¡¯ particles
such as hyperons, pion condensates, kaon condensates, quarks and etc. Here $\rho_{tr}$ is the
transition density of singular particles. To date, the value of $\rho_{tr}$
is uncertain. For example, Tsuruta et al.(2009) (see Ref.~\refcite{Tsuruta09})
gave an estimate $\rho_{tr}\sim 4\rho_{0}$.

What we are most interested in is how a SMF can influence
Landau levels of degenerate and relativistic electrons and their Fermi energy. Many
authors (see Ref.~\refcite{Das12},~\refcite{Dong13},~\refcite{Gao15},~\refcite{Ferrer15}) have carried out detailed
studies on the influences of a SMF on the composition
and the EOS of a NS. According to the requirement of quantization of
Landau levels, we introduced the Dirac$-\delta$ function (see Ref.~\refcite{Gao11a},~\refcite{Gao12a}),
and obtained a particular solution to $E_{\rm F}(e)$,
\begin{equation}
E_{\rm F}(e)\simeq 43.44\times \left(\frac{\rho}{\rho_{0}}\frac{Y_e}{0.0535}\frac{B}{B_{\rm cr}}\right)^{1/4}~~
~\rm MeV~~(B\gg B_{\rm cr}),
\label{1}
\end{equation}
where $Y_{e}$ is the electron fraction, which
is defined as $Y_{e}=n_{e}/n_{B}$, where $n_{e}$ and $n_{B}$ are the electron
number density, and the baryon number density, respectively (see Ref.~\refcite{Gao11a},~\refcite{Gao12a}).
Furthermore, we deduced a general expression for $P_{e}$, the pressure of relativistic
electrons (see Ref.~\refcite{Gao13}), discussed the quantization of the electron Landau levels,
and explored the influence of quantum electrodynamics effects on the EoS. The
main conclusions included: The higher the magnetic field intensity, the bigger
the electron pressure, and the high pressure is caused by high Fermi energy of
electrons; the total pressure of a magnetar is always anisotropic; compared with an
ordinary radio pulsar, a magnestar might be a denser NS if the anisotropic
total pressure is taken into consideration; a magnestar might have larger mass if the positive energy
 contribution of the magnetic field energy to the EoS is taken into consideration (see Ref.~\refcite{Gao13}).

Our research results pose a challenge to the prevalent viewpoint (see Ref.~\refcite{Lai91},~\refcite{Harding06}): In a
 SMF, the higher the magnetic field intensity $B$, the lower the
 electron Fermi energy and the electron pressure. This prevalent viewpoint essentially goes against
 the real requirement of the quantization of landau levels, due to the introduction of
 an artificial and false assumption and the application of the solution of a non-relativistic
 electron cyclotron motion equation (see Ref.~\refcite{Gao13}) for specific information).

 Recently, after a careful examination, we found some inadequacies of our theoretical model,
 mainly including the following aspects: 1) No consideration was given to the stability
 of Landau levels of electrons in a strong magnetic field. Till now, there has yet
 been no any relevant works or explicit analytical expression on the
 stability coefficient $g_{n}$ in the physics community due to the
 complexity of this issue; 2) No explicit analytic expression for $E_{\rm F}(e)$ and
 $n_{e}$ was provided. There was no comparison between the relationship of $E_{\rm F}(e)$ and
 $n_{e}$ in a SMF with that in a weak magnetic field approximation ($B^{*}\ll 1$), based
 on which, the variation range of the magnetic field index $\beta$ in the expression is defined;
 3) In the expression of $E_{\rm F}(e)$, the application scope of $B$ was not clearly defined because the
 Fermi surfaces (in the momentum space) of electrons in a non-relativistic magnetic field are
 basically spherically symmetrical, whereas the Fermi ball (in the momentum space) is turned into
 Landau cylinder in a relativistic magnetic field (see Ref.~\refcite{Landau65},~\refcite{Canuto77},~\refcite{Peng07}); 4) The most important thing is that the physical meaning of the magnetic field index $\beta$ ($\beta=1/4$) in
 the expression of $E_{\rm F}(e)$ (see Ref.~\refcite{Gao11a},~\refcite{Gao12a}) is not clear.

 With the increase in $B$, the Landau cylinder becomes longer
 and narrower. When the SMF is too high, the Landau cylindrical
 space will be streamlined into a one-dimensional linear chain, making our model no
 longer applicable. Simply speaking, much detailed information in our previous works
 has been neglected (especially, ignoring the discrepancy of different Landau levels of
 electrons) in the process of derivation of $E_{\rm F}(e)$ and/or $P_{e}$. Therefore, it is
 of great importance to modify the expression $E_{\rm F}(e)$ in a SMF.

This paper is organized as follows: In Section 2, we review the relationship
between $E_{\rm F}(e)$ and $n_{e}$ in the weak magnetic field approximation; in
Section 3, we deduce a general expression of $E_{\rm F}(e)$ in a SMF by introducing
the stability coefficient of the Landau levels, and modify
the particular solution to $E_{\rm F}(e)$, and in Section 4, we present our summary
and discussion.
\section{The Fermi Energy in The Weak Magnetic Field Approximation}
This part mainly refers to Ref.~\refcite{Li16}. Based on the basic
definition of the Fermi energy of relativistic electrons,
we obtain a particular solution to $E_{\rm F}(e)$,
\begin{equation}
E_{\rm F}(e)=60\times(\frac{\rho}{\rho_{0}})^{1/3}(\frac{Y_{e}}{0.005647})^{1/3}~~~~({\rm MeV}).
\label{2}
\end{equation}
which is suitable for relativistic electron matter region in a NS. By means of
numerical simulation, we obtained some analytic expressions of $Y_{e}$ and $\rho$
for several reliable EoSs with which we can estimate $E_{\rm F}(e)$ at
any matter density by combining these analytical
expressions with boundary conditions\cite{Li16}.

Although $E_{\rm F}(e)$ in a weak magnetic field approximation could be
expressed as the function of $Y_{e}$ and $\rho$, the Fermi energy of electrons
is solely determined by the electron number density $n_e$. Since electrons are
extremely relativistic, the dimensionless electron Fermi momentum $x_{e}=p_{\rm F}(e)/m_{e}c\gg 1$,
then we obtained the relationship between $E_{\rm F}(e)$ and $n_e$,
\begin{eqnarray}
E_{\rm F}(e)&&= m_{e}c^{2}(1+ x_{e}^{2})^{1/2}\approx m_{e}c^{2}x_{e}\nonumber\\
&&=m_{e}c^{2}(n_{e}3\pi^{2}\lambda_{e}^{3})^{1/3}\nonumber\\
 &&=\hbar c(3\pi^{2}n_{e})^{1/3}=6.12\times 10^{-11}n_{e}^{1/3}~({\rm MeV}),
 \label{3}
\end{eqnarray}
in a weak magnetic field approximation, where~$\lambda_{e}=h/m_{e}c=
2.4263\times 10^{-10}$~cm is the electron Compton wavelength.

All the other Fermi parameters of electrons are also solely determined by the
number density of free electron gas, $n_{e}$. For example, the electron Fermi
momentum $p_{\rm F}(e)=\hbar k_{\rm F}= \hbar(3\pi^{2}n_{e})^{1/3}$, where $k_{\rm F}=
(3\pi^{2}n_{e})^{1/3}$ is the Fermi wave vector of electrons. For the Fermi
kinetic energy of relativistic electrons, $E_{K}^{\rm F}(e)\approx cp_{\rm F}(e)= c\hbar(3\pi^{2}
n_{e})^{1/3}$, and $E_{K}^{\rm F}(e)\gg m_{e}c^2$). However, the relations between
$n_{e}$ and $\rho$ in different density regions of a NS are usually unknown, and the
known relations of $n_{e}$ and $\rho$ depend on the EoS in some specific matter models,
and on the analytical expression of $Y_e$ and $\rho$ obtained from the EOS in a certain matter model.
\section{Electron Fermi Energy in a Superhigh Magnetic Field}\label{III}
\subsection{Stability of Electron Landau Level}
We now consider a uniform magnetic field $B$ directed
along the z-axis. In this case, in the Landau gauge
the vector potential $\vec{A}$ reads $\vec{A}=(-B_{y}, 0, 0)$. For
extremely strong magnetic fields, the cyclotron energy
becomes comparable to the electron rest-mass energy,
and the transverse motion of the electron becomes relativistic.
A relativistic magnetic field is often called the quantum critical magnetic field ($B_{\rm cr} =
m_{e}^{2}c^{3}/e\hbar=4.414\times 10^{13}$ Gauss), which is obtained from the
relation $\hbar\omega=m_{e}c^2$.  The electron energy levels may be
obtained by solving the relativistic Dirac equation in a
strong magnetic field with the result
\begin{equation}
E_{e}^{2} = m^{2}_{e}c^{4}(1+2\nu\frac{B}{B_{cr}})+p^{2}_{z}(e)c^{2},
\label{4}
\end{equation}
where the quantum number $\nu$ is given by $\nu=n +1/2+
\sigma^{'}$, the Landau level number $n = 0, 1, 2, \cdots $,
spin $\sigma^{'}=\pm 1/2$, and the quantity $p_{z}(e)$ is the
$z$-component of the electron momentum, deemed as a continuous
function. In a SMF, the maximum $z$-momentum of
electrons $p^{F}_{z}(e)$ is defined (see Ref.~\refcite{Lai91}) by
\begin{equation}
 (p^{F}_{z}(n)c)^{2}+ m^{2}_{e}c^{4}+(2n+1+\sigma)m^{2}_{e}c^{4}B^{*} \equiv E^{2}_{F}(e),
 \label{5}
 \end{equation}
where the range of $p^{F}_{z}(e)$ is $0\sim E_{\rm F}(e)/c$.
For given values of the magnetic field intensity, the Fermi energy and the
$z$-momentum of electrons, the electron Landau level
number $n$ is given by
\begin{equation}
 n(\sigma= -1)= Int\left[\frac{1}{2B^{*}}[(\frac{E_{\rm F}(e)}{m_{e}c^{2}})
 ^{2}-1-(\frac{p^{F}_{z}(e)}{m_{e}c})^{2}]\right],
 \label{6}
\end{equation}
\begin{equation}
 n(\sigma=1)= Int\left[\frac{1}{2B^{*}}[(\frac{E_{\rm F}(e)}{m_{e}c^{2}})
 ^{2}-1-(\frac{p^{F}_{z}(e)}{m_{e}c})^{2}]-1\right],
     \label{7}
\end{equation}
where $Int[x]$ denotes an integer value of the argument $x$,
and $\sigma=2\sigma^{'}=\pm 1$ is the spin projection value.

In a weak magnetic field $B^{*}\ll 1$, for the electron gas in the
non-degenerate limit (temperature different from zero), the maximum
Landau level number, $n_{m}\rightarrow \infty$. However, the maximum
Landau level number $n_{m}$ is set by the condition $p^{F}_{z}(e)\geq 0$ or
\begin{equation}
 E^{2}_{F}(e)\geq m^{2}_{e}c^{4}(1+2\nu\frac{B}{B_{cr}}),
\label{8}
\end{equation}
for highly degenerate electron gas in a SMF (see Ref.~\refcite{Lai91}).
It is obvious that the maximum of the electron Landau level number decreases
with $B$ when $E_{\rm F}(e)$ and $p^{F}_{z}(e)$ are given. This is because the
higher the magnetic field intensity, the more unstable the Landau
levels of electrons, and the bigger the Landau level number $n$, the lower the
Landau-level stability.

Indeed, the issue concerning the Landau-level stability of charged particles
in a SMF is so complicated that there has not been any
relevant work or explicit analytical expression in the physics community. In
our previous works (see Ref.~\refcite{Gao11a},~\refcite{Gao12a}), we did not take into consideration the
Landau-level stability of electrons in a SMF that limits
the application of our model. In this work, a new quantity, $g_n$, the Landau-level
stability coefficient of electrons in a SMF, is introduced. Considering the uncertainty of
the electron microscopic states in a SMF, we assume that $g_n$ takes the power-law form
\begin{equation}
g_{n} = g_{0}n^{\alpha}~~~~~(n\geq 1), \label{9}
\end{equation}
where $n$, $g_{0}$ and $\alpha$ are the Landau level number, the ground-
state level stability coefficient, and the stability index of Landau levels,
respectively.  When $n=1$, $g_{1}=g_{0}$, i.e., the ground-state
level has the same stability as that of the first excited level.  According to
quantum mechanics, the electrons at a higher energy level are prone to have
excited transitions towards a lower energy level. The bigger the Landau level
number, the shorter the level-occupying time for electrons, and the lower the
Landau-level stability, the higher the probability of the excited transition.

Since the ground state level has the highest stability and $g_{n}$ decreases
with increasing $n$, i.e., $g_{n}<g_{n-1}<g_{n-2}$, the stability index $\alpha$
should be negative. The main reasons are as follows: If $\alpha=0$, then $g_{n}=
g_{n-1}=\cdots =g_{1}=g_{0}$, i.e., all the Landau levels have the same stability,
and the maximum of the Landau level number, $n_{m}$, can take any high value. This
scenario is essentially corresponding to a weak magnetic field approximation; if
$\alpha>0$, then $g_{n}>g_{n-1}>\cdots=g_{1}=g_{0}$, and under such a condition, a
higher Landau level number means a higher stability, and $n$ can also take any high
value, which is clearly contrary to the principles of quantum mechanics. According to
the analysis above, for degenerate and relativistic electrons in a SMF, the
Landau-level stability index, $\alpha<0$.  Based on Eq.(9), we make a
schematic diagram of $g_{n}$ and $\alpha$, shown in Fig. 1.
\begin{figure*}[htb]
\centerline{\psfig{file=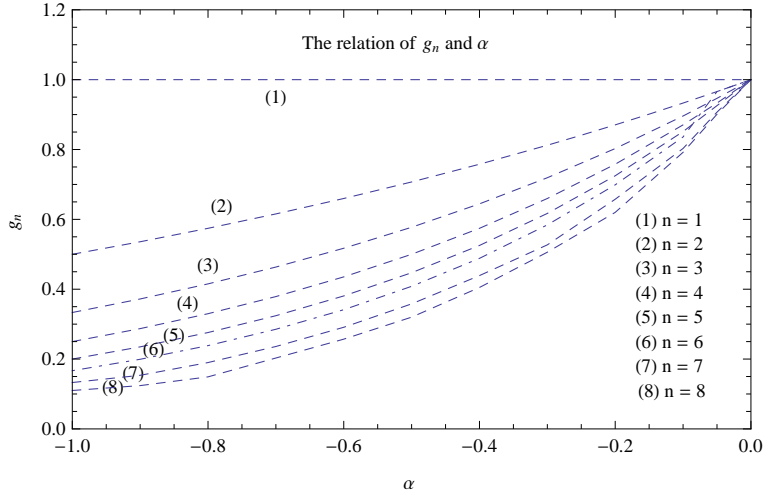,width=4.0in}}
\vspace*{8pt}
\caption{The diagrams of $g_{n}$
vs. $\alpha$ for electrons in a superhigh magnetic field.
\protect\label{fig1}}
\end{figure*}

As seen from Fig.1, for a given Landau level with $n\geq 1$, the coefficient $g_{n}$
increases with $\alpha$ slowly, and the bigger the Landau level number $n$, the faster the change of $g_{n}$
with $\alpha$. From  discussions above, the bigger the Landau level
number $n$, the greater the influence of the stability index $\alpha$ on
$g_{n}$, and the larger the probability of a particle's transition from a
higher energy level into a lower energy level.

It should be pointed out that, in atomic physics and statistics
mechanics (see Ref.~\refcite{Kubo65},~\refcite{Pathria03}) the statistical weight describes the
energy state density of microscopic particles, in other words, a higher energy level
number means a bigger statistical weight. The higher the quantum number $\nu$,
the wider the energy level width, and the more the microscopic state number of particles is.
Here the Landau-level stability coefficient for electrons and the statistical
weight are two totally different concepts.
\subsection{The Energy State Density of Electrons in Phase Space }
In a SMF, the Fermi surface of electrons
becomes a narrow Landau cylinder.  Combining $B_{\rm cr}=m^{2}_{e}c^{3}/
e\hbar$ with $\mu_{e}= e\hbar/2m_{e}c$, Eq.(4) is modified as
\begin{equation}
  E_{e}^{2}= m_{e}^{2}c^{4}+p_{z}^{2}(e)c^{2}
 +(2n+1+\sigma)2m_{e}c^{2}\mu_{e}B,~
 ~\label{10}
\end{equation}
where $\mu_e\sim 0.927\times10^{-20}$ ergs~Gauss$^{-1}$ is the
magnetic moment of an electron. The continuous physical variables $p_{x}$
and $p_{y}$ (see Ref.~\refcite{Landau65}), adopted in a non-relativistic magnetic field,
will be no longer applicable. Thus, a quantized or discrete relativistic
variable $p_{\perp}$ must be adopted for replacement, where $p_{\perp}$ is
the electron  momentum perpendicular to the magnetic field, $p_{\perp}=
m_{e}c((2n+1+\sigma)B^{*})^{1/2}$. The microscopic state number in a
6-dimension phase-space element $dxdydzdp_{x}dp_{y}dp_{z}$ is $dxdydzdp_{x}dp_{y}dp_{z}/h^{3}$.
By using the relation $2\mu_{e}B_{\rm cr}/m_{e}c^{2}= 1$ and summing
over the electron energy states in a 6-dimension phase space, we can express
the electron energy state density  $N_{pha}$ as follows
\begin{eqnarray}
N_{pha}&&=\frac{2\pi}{h^{3}}\int dp_{z}\sum_{n=0}^{n_{m}
 (\sigma,B^{*})}\sum g_{n}\times \nonumber\\
 &&\int \delta(\frac{p_{\perp}}{m_{e}c}-[(2n+1+\sigma)B^{*} ]
 ^{1/2}) p_{\perp}dp_{\perp}.
 \label{11}
 \end{eqnarray}
where $\delta(p_{\perp}/m_{e}c-[(2n+1+\sigma)B^{*}]^{1/2})$ is the
Dirac-$\delta$ function for electrons in a SMF.
The physical significance of the Dirac-$\delta$ function lies in that,
there doesn't exist any microscopic quantum state between the $n$-th
and $n+1$-th Landau torus due to the strong quantization of the electron
Landau levels.  For the ground state level, the electron spin is
antiparallel to $B$, so the Landau level is non-degenerate ($n=0, \sigma=-1$);
whereas higher levels are doubly degenerate ($n\geq 1, \sigma=\pm 1$).
Thus Eq.(11) can be rewritten as
\begin{eqnarray}
 N_{pha}&&= 2\pi(\frac{m_{e}c}{h})^{3}\int_{0}^{\frac{p^{F}_{z}(e)}{m_{e}c}} d(\frac{p_{z}}
 {m_{e}c})g_{n}\times \nonumber\\
 &&[\sum_{n = 0}^{n_{m}(B^{*},\sigma=-1)}\int\delta(\frac{p_{\perp}}{m_{e}c}-(2nB^{*} )^{1/2})
 (\frac{p_{\perp}}{m_{e}c})d(\frac{p_{\perp}}{m_{e}c})\nonumber\\
 &&+\sum_{n=1}^{n_{m}(B^{*},\sigma=1)}\int \delta(\frac{p_{\perp}}{m_{e}c}-(2(n+1)B^{*})^{1/2})
 (\frac{p_{\perp}}{m_{e}c})d(\frac{p_{\perp}}{m_{e}c})],\label{12}
 \end{eqnarray}
 The upper limit of summation on Eq.(12) is $n_{m}(B^{*})$, which has the following approximate relation,
\begin{eqnarray}
 &&n_{m}(B^{*},\sigma=1)\simeq n_{m}(B^{*},\sigma=-1) \nonumber\\
 &&= n_{m}(B^{*})\approx Int\left[\frac{1}{2B^{*}}\times (\frac{E_{\rm F}(e)}{m_{e}c^{2}})^{2}\right].
 \label{13}
 \end{eqnarray}
when $n_{m}(B^{*})\gg 1$. In deriving the above expression,
we have taken into account the dependence of $n_{m}(B^{*})$ on
$p^{F}_{z}(e)\geq 0$, and assumed $\frac{E_{\rm F}(e)}{m_{e}c^{2}}\gg 1$,
and the lowest limit $p^{F}_{z}(e)\rightarrow 0$. Inserting Eq.(13) into Eq.(12) yields
\begin{eqnarray}
N_{pha}&&=2\pi(\frac{m_{e}c}{h})^{3}g_{0}\int_{0}^{\frac{p^{F}_{z}(e)}{m_{e}c}}\sqrt{2B^{*}}\times \nonumber\\
&&\sum_{n= 0}^{n_{m}(B^{*})}n^{\alpha}(\sqrt{n}+\sqrt{n+1})d(\frac{p_{z}}{m_{e}c}).
\label{14}
 \end{eqnarray}
With a more rigorous replacement of integral upper limit $\int_{0}
^{\frac{p^{F}_{z}(e)}{m_{e}c}}\rightarrow\int_{0}^{\frac{E_{\rm F}(e)}
{m_{e}c^2}}$ (the range of $p^{F}_{z}(e)$ is $0\sim E_{\rm F}(e)/c$),
Eq.(14) can be simplified as,
 \begin{equation}
N_{pha} = 2^{\frac{5}{2}}\pi \sqrt{B^{*}}(\frac{m_{e}c}{h})^{3}g_{0}
\int_{0}^{\frac{E_{\rm F}(e)}{m_{e}c^(2)}}\sum_{n=0}^{n_{m}(B^{*})}n^{\alpha+1/2}d(\frac{p_{z}}{m_{e}c}).
\label{15}
 \end{equation}
Note that if the matter density is so high that the electron
longitudinal kinetic energy exceeds its rest-mass energy, or
if the magnetic field is so high that the electron cyclotron
energy exceeds its rest-mass energy, then the electron becomes relativistic.
Here we introduce a ratio $q(\alpha)$, which is defined as $q(\alpha)=
I_{1}/I_{2}$, $I_{1}=\int_{0}^{n_{m}(B^{*})} n^{\alpha+1/2}dn$ and $I_{2}=\sum_{n=0}
^{n_{m}(B^{*})}n^{\alpha+1/2}$.  For a given index $\alpha$ ($\alpha<0$),
assuming $n_{m}(B^{*})$ to be 6, 8, 10, 15, 20 and 30 at random, we can
calculate the corresponding values of $q(\alpha)$, as listed in Table 1.
\begin{table*}[htb]
\tbl{Values of $q(\alpha)$ assuming different $\alpha$ and \textbf{$n_{m}(B^{*})$}. }
{\begin{tabular}{@{}ccccc@{}} \toprule
$n_{m}(B^{*})$ & $q(\alpha)$& $q(\alpha)$ &$q(\alpha)$ &$q(\alpha)$   \\
   & $(\alpha=-0.1)$& $(\alpha=-0.5)$ &$(\alpha=-0.8)$ &$(\alpha=-0.95)$   \\
\colrule
6 &0.83 &0.86 &0.88 & 0.90\\
8 &0.87 &0.89 & 0.91&0.93 \\
10&0.89 &0.91 &0.93 & 0.94\\
 15 &0.92 &0.94 &0.95 &0.96 \\
20 & 0.94&0.95 & 0.97 & 0.97\\
25 &0.95 &0.96 &0.97 & 0.97 \\
30 &0.96 &0.97 &0.97 & 0.98\\
\botrule
\end{tabular}\label{ta1}}
\end{table*}

From Table 1, it is easy to see that $q(\alpha)$ increases
with $n_{m}(B^{*})$, and $q\simeq 1$ if $n\gg 1$.  Thus, the
following summation formula is approximately replaced by an integral equation,
\begin{equation}
\sum_{n=0}^{n_{m}(B^{*})}n^{\alpha+\frac{1}{2}}\simeq \int_{0}^{n_{m}(B^{*})}n^{\alpha+\frac{1}{2}}dn
 = \frac{2}{2\alpha+3}n^{\alpha+\frac{3}{2}}_{m},
 \label{16}
\end{equation}
when $n_{m}(B^{*})\geq$ 6. Then Eq.(15) can be rewritten as
\begin{equation}
 N_{pha}= \frac{2^{\frac{7}{2}}}{2\alpha+3}\pi \sqrt{B^{*}}
 (\frac{m_{e}c}{h})^{3}g_{0}\int_{0}^{\frac{E_{\rm F}(e)}{m_{e}c^(2)}}\left[\frac{1}{2B^{*}}(\frac{E_{F}(e)}{m_{e}c^{2}})^{2}\right]^{\alpha+3/2}
 d(\frac{p_{z}}{m_{e}c}).
   \label{17}
 \end{equation}
\subsection{The Fermi Energy of Electrons in A Superhigh Magnetic Field }
Since $d(\frac{p_{z}}{m_{e}c})=d(\frac{p_{z}c}{m_{e}c^{2}})$,
Eq.(17) can be further simplified
  \begin{equation}
N_{pha}=\frac{2^{2(1-\alpha)}}{2\alpha+3}\frac{\pi}{(B^{*})^{1+\alpha}}g_{0}(\frac{m_{e}c}{h})^{3}
(\frac{E_{F}(e)}{m_{e}c^{2}})^{(2\alpha+4)}.
\label{18}
\end{equation}
In order to modify the formula for $E_{\rm F}(e)$ in a SMF,
let us refer to the Pauli exclusion principle (PEP) (see Ref.~\refcite{Pauli25}). According to
the PEP, there are no two identical fermions occupying the same quantum state
simultaneously, thus highly degenerate electrons in a SMF
have to fill quantum states from the lowest Landau level (the ground-state level) to
the highest Landau level. In the mean time, according to quantum mechanics,
the larger the electron Landau level number $n$, the more unstable the electron
Landau level, and the electrons at a higher energy level are prone to have excited
transitions towards a lower energy level by losing energy (e.g., releasing photons).

The electron Fermi energy, as the highest ocupied state energy, corresponds to the maximum electron Fermi momentum. The
resulted electron degeneracy pressure contributes a small fraction of
the total dynamic pressure (mainly from neutron degeneracy pressure)
countering against gravity collapse of a NS.  Thus, the PEP not only
explains higher energy levels of electrons, but also is responsible for
the stability of the NS matter. In a NS, the electron number
density $n_{e}$ is determined by
\begin{equation}
 n_{e} =  N_{A}\rho Y_{e},
 \label{19}
\end{equation}
where $N_{A}=6.02\times 10^{23}$ is the Avogadro constant. According
to  PEP, the electron energy state number equals the electron number in a unit volume, we get
\begin{eqnarray}
N_{pha}&&=\frac{2^{2(1-\alpha)}}{2\alpha +3}\frac{\pi}{(B^{*})^{1+\alpha}}g_{0}(\frac{m_{e}c}{h})^{3} \nonumber\\
&&\times(\frac{E_{F}(e)}{m_{e}c^{2}})^{2\alpha +4} = N_{A}\rho Y_{e}=n_{e}.
\label{20}
\end{eqnarray}
By solving Eq.(20), we obtain
\begin{eqnarray}
E_{F}(e)&&=\left(\frac{2\alpha +3}{2^{(2-\alpha)}\pi g_{0}}\right)^{\frac{1}{2(\alpha+2)}}\times \nonumber\\
&&(\frac{h}{m_{e}c})^{\frac{3}{2(\alpha+2)}}  
 m_{e}c^{2}\left(B^{*}\right)^{\frac{1+\alpha}{2(\alpha+2)}}n_{e}^{\frac{1}{2(\alpha+2)}}.
\label{21}
\end{eqnarray}
Eq.(21) is a new general expression of $E_{\rm F}(e)$ in a SMF, where $\frac{1+\alpha}{2(\alpha+2)}$ describes the
magnetic field index of the expression. For the sake of convenience, the magnetic field index
is denoted by a quantity $\beta$, $\beta=\frac{1+\alpha}{2(\alpha+2)}$.

In order to discuss a reasonable range of $\alpha$, we generate a schematic diagram of $\beta$ and $\alpha$.
\begin{figure*}[htb]
\centerline{\psfig{file=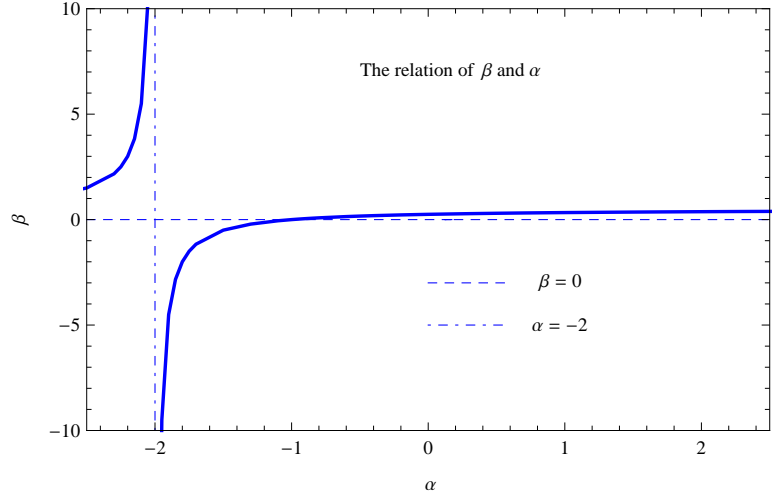,width=4.0in}}
\vspace*{8pt}
\caption{The relation between $\beta$ and $\alpha$..
\protect\label{fig2}}
\end{figure*}
In Fig. 2, the dot-dashed line represents a singular point of $\alpha=-2$;
the dashed line represents $\beta=0$, corresponding to $\alpha=-1$. The
physical significance of $\alpha=-1$ is that $E_{\rm F}(e)$ does not change
with the variation of $B$, which equivalents to the case of weak-magnetic field
approximation. The reasonable range of $\alpha$ is thus estimated as $\alpha<0$ but
$\alpha\neq -1, -2$. From Fig. 2, one can easily judge the relationship between
$E_{\rm F}(e)$ and $B$: When $\alpha <-2$ or $-1< \alpha <0$, the magnetic field
index $\beta>0$, and $E_{\rm F}(e)$ increases with $B$; when $-2< \alpha <-1$,
the magnetic field index $\beta<0$, and $E_{\rm F}(e)$ decreases with $B$.

Whether $\beta$ is a positive or negative number depends on actual value of $\alpha$.
From the relation of $E_{\rm F}(e)$ and $n_e$ in a weak magnetic field
approximation, it can be seen that $E_{\rm F}(e)$ is solely determined by $n_e$,
and $E_{\rm F}(e)\propto n_{e}^{1/3}$ (refer to Eq.(3) in Section 2).
From the general expression of $E_{\rm F}(e)$ in a SMF (Eq.(20)),
we can see that $E_{\rm F}(e) \propto \left(B^{*}\right)^{\frac{1+\alpha}{2(\alpha+2)}}
n_{e}^{\frac{1}{2(\alpha+2)}}$, i.e., $E_{\rm F}(e)$ bears the
relation not only to $n_e$ but also to $B$, and the latter has great influence
on the former by modifying electron phase space. It is also worth noticing
that, the dimension of $E_{\rm F}(e)$ always remains unchanged no matter in a
weak magnetic field or in a SMF, i.e., $E_{\rm F}(e)$ is
proportional to $n_{e}^{1/3}$.
Thus, we obtain
\begin{equation}
\frac{1}{2(\alpha+2)}=\frac{1}{3},
 \label{22}
\end{equation}
by comparing Eq.(3) with Eq.(21).
Solving Eq.(22) yields $\alpha=-0.5$, which lies in
a reasonable range of $\alpha$, as estimated above. Inserting $\alpha=-0.5$
into Eq.(22), we get the magnetic field index $\beta=1/6$.
Compared with Eq.(1), the value of $\beta$ obtained in this paper
decreases by $1/12$. In spite of the minimum disparity, the modified magnetic field
index is obviously superior to that of previous one (see Ref.~\refcite{Gao11a},~\refcite{Gao12a}).
This is because, in our previous works (see Ref.~\refcite{Gao11a},~\refcite{Gao12a}), the magnetic field index
$\beta$ ($\beta=1/4$) corresponds to $\alpha=0$, which means that different Landau
levels have the same stability, and the differences between two different
Landau levels are neglected.

In order to obtain an analytic expression for $E_{\rm F}(e)$ in a SMF, we assume that
the ground state level has the highest stability, and the maximum value of $g_n$ is $g_0= 1$.
This provides us with much convenience for theoretical derivations.
 Then the electron Landau-level stability coefficient (see Eq.(7)) can be
expressed as $g_1=g_0=1, (n=0, 1)$ and $g_n=n^{-\frac{1}{2}}~~(n\geq 2)$.

Inserting $\alpha=-0.5$ into Eq.(21), we have
\begin{eqnarray}
E_{\rm F}(e)&&=5.84\times 10^{-11}(n_{e}^{'})^{1/3} \nonumber\\
&&=5.84\times 10^{-11}(B^{*})^{1/6}n_{e}^{1/3} \nonumber\\
&&=5.84\times 10^{-11}(\frac{B}{B_{cr}})^{1/6}n_{e}^{1/3}~(\rm~MeV),
 \label{23}
\end{eqnarray}
where $n_{e}^{'}=n_{e}(B^{*})^{1+\alpha}=n_{e}(B^{*})^{1/2}$  after considering the magnetic effects.
Eq.(23) is a newly modified general expression for $E_{\rm F}(e)$ and $n_e$ in a SMF.

By modifying the phase space of relativistic electrons, a SMF can enhance $n_e$, and decrease the maximum of electron Landau level number, resulting in a redistribution of electrons. As mentioned above, the strongly degenerate electrons have to occupy all possible microscopic states up to the highest Landau level, due to the requirement of the PEP. The enhanced $n_e$ in a SMF means an increase in $E_{\rm F}(e)$, corresponding to an increase in the electron degeneracy pressure.

If $B$ is too high, e.g., $B> 10^{17}$ Gauss, the Landau cylinder will
continuously elongate in the direction of $B$, and could become a narrow electron
chain, then the maximum Landau level number is estimated to be $n_{m}=1$ or 2.
Under such circumstances, the premise of theoretical derivations in this paper will no longer apply.
Thus the applicable conditions for Eq.(23) are limited by $\rho\geq 10^{7}$~g cm$^{-3}$ and $B_{\rm cr}\ll B\leq 10^{17}$~Gauss.

The Fermi energy of electrons in a magnetar is related to $Y_e$, $\rho$ and $B$.
Inserting $n_{e}=N_{A}Y_{e}\rho$ into Eq.(23), we obtain a newl solution to $E_{\rm F}(e)$ in a SMF,
\begin{eqnarray}
E_{\rm F}(e)&&=59.1\times \left(\frac{Y_e}{0.005647}\right)^{1/3}\left(\frac{\rho}{\rho_{0}}\right)^{1/3}\left(B^{*}\right)^{1/6}  \nonumber\\
&&=59.1\times \left(\frac{Y_e}{0.005647}\right)^{1/3}\left(\frac{\rho}{\rho_{0}}\right)^{1/3}\left(\frac{B}{B_{\rm cr}}\right)^{1/6}~(\rm MeV),
\label{24}
\end{eqnarray}
where the relation of $Y_{e}=Y_{p}\approx \frac{n_{e}}{n_{n}}\approx
0.005647\times(\frac{\rho}{\rho_{0}})$ (see Eq.(12) of Ref.~\refcite{Li16}) is used.
We make schematic diagrams of $E_{\rm F}(e)$
and $\rho$ in different magnetic fields, as shown in Fig. 3
\begin{figure*}[htb]
\centerline{\psfig{file=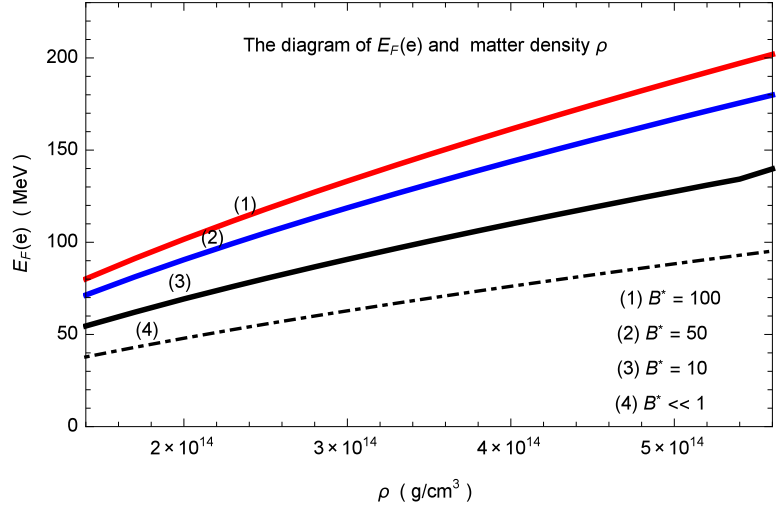,width=4.0in}}
\vspace*{8pt}
\caption{The relations of $E_{\rm F}(e)$ and $\rho$ in different magnetic fields.
The matter density ranges $\rho \sim (1.4\times
10^{14}-5.6\times 10^{14})$~g cm$^{-3}$.  The solid lines represent different
strong magnetic fields and the dot-dashed line represents the weak magnetic
field approximation ($B^{*}\ll 1$).
\protect\label{fig3}}
\end{figure*}
In Fig. 3 the curves 1, 2 and 3 are fitted by Eq.(24), and the curve 4 is
fitted by Eq.(2). When the matter density remains constant, $\rho=\rho_0$,
we calculate the values of $E_{\rm F}(e)$ to be
$(102.41-162.58)$, $(91.24- 144.84)$ and $(69.77 -110.75)$ MeV, respectively when
$B^{*}=100$, $50$, and $10$, respectively.

In order to compare the modified particular solution to $E_{\rm F}(e)$ with that in
our previous works (see Ref.~\refcite{Gao12a}), we make schematic diagrams of
$E_{\rm F}(e)$ and $B$, as shown in Fig.4.

\begin{figure*}[htb]
\centerline{\psfig{file=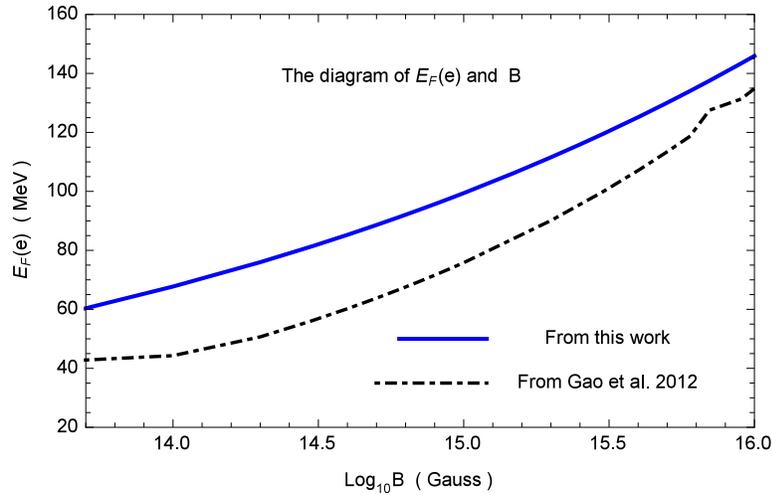,width=4.0in}}
\vspace*{8pt}
\caption{The relations of $E_{\rm F}(e)$ and $B$ in SMFs.
The range of $B$ is $(5.0\times 10^{13}-1.0\times 10^{16})$~Gauss. In order to
facilitate the calculation, the matter density is assumed as
$\rho=\rho_0$, arbitrarily.  The solid line and the dot-dashed line
are fitted by Eq.(24) and Eq.(1), respectively.
\protect\label{fig4}}
\end{figure*}
When $\rho=\rho_0$, and $B\sim (5.0\times 10^{13}-1.0\times 10^{16})$~Gauss,
we calculate the values of $E_{\rm F}(e)$ to be $(60.34-145.92)$ MeV, and
$(46.1-134.83)$ MeV, respectively, by using Eq.(24) and Eq.(1), respectively.
Fig. 4 clearly shows that the value of $E_{\rm F}(e)$ obtained with Eq.(24)
is slightly bigger than that obtained with Eq.(2). This is mainly because
that both the Fermi energy coefficient (59.1 MeV) and the density index
(1/3) in Eq.(24) are higher than those in Eq.(1), although the magnetic field index
$\beta$ in Eq.(24) is lower than that in Eq.(2).
\section{Summary and Discussion}\label{IV}
By introducing the stability coefficient of the electron Landau levels,
we re-derive the general expression of $E_{\rm F}(e)$ in a
SMF, and obtained a modified particular
solution to $E_{\rm F}(e)$. The solution has a lower magnetic
field index of $\beta=1/6$, which is lower than the previous
one by a factor of $1/12$.  Since there exists the discrepancy of
stability in different Landau levels, we believe that this solution
to $E_{\rm F}(e)$ is superior to the previous one (see Ref.~\refcite{Gao11a},~\refcite{Gao12a}).

Just like in a weak magnetic field approximation, the value of the
electron Fermi energy is determined solely by the electron
number density. A SMF modifies the phase space
of electrons, and thus increases the electron number density. The possible
reasons are as follows:

(I) According to the requirement of electrical neutrality,
the simple neutron decay and continuous electron capture occur simultaneously
in the NS interior. However, a SMF will cause the
former process to be faster than the latter (see Ref.~\refcite{Chakrabarty97}). Since more
neutrons in a SMF will be converted into protons, the
proton fraction $Y_p$ ($Y_p= Y_e$) will increases, and the electron
number density ($n_{e}=N_{A}Y_{e}\rho$) will also increase correspondingly.

(II) According to nuclear physics, the proton fraction describes the
asymmetry of nuclear matter, and the value of $Y_p$ bears close relation to the
symmetry energy, the symmetrical energy gradient, the incompressible
coefficient, the volume bound energy and other parameters of nuclear
matter (see Ref.~\refcite{Wang14},~\refcite{Wang15}).  A superhigh magnetic field may increase
the asymmetry of nuclear matter and improve the proton fraction. Thus, the
average electron density of nuclear matter also increases correspondingly.

Although the structure, properties and EoS of nuclear matter are
strongly influenced by a SMF, there has been  no
relevant and detailed research results in the physics community, the study
on how a SMF influences the asymmetry (the influence
on the proton abundance in particular) of nuclear matter will become
one of our future research directions. Since the electron Fermi energy is one of most important parameters of
EoS, a SMF will influence the
EoS of a NS, as well as on the electron Fermi energy. Meanwhile, the particular
solution to the electron Fermi energy obtained in this paper will surely
influence calculations of the neutron-decay rates, the electron capture
rates and the soft X-ray luminosity of a magnetar (see Ref.~\refcite{Gao11a},~\refcite{Gao11b},~\refcite{Gao12a},~\refcite{Gao12b}).

As known to all, the cooling processes of a NS can be
categorized into direct Urca and modified Urca reactions (see Ref.~\refcite{Yakovlev01}).
Theoretical research shows that, direct Urca reactions in a ordinary NS with $B\ll B_{\rm cr}$ are strongly
suppressed by Pauli blocking in a system composed of neutrons,
protons and electrons, due to a high threshold (see Ref.~\refcite{Boguta77},~\refcite{Boguta81},~\refcite{Boguta83},~\refcite{Takahashi08},~\refcite{Henley07}) for the proton fraction $Y_{p}\geq $1/9. Direct Urca reactions could take place (see Ref.~\refcite{Baiko99},~\refcite{Lai91},~\refcite{Yakovlev01}) in the core of a supermassive NS, where $Y_{p}$ could be
higher than 0.11. However, when in a SMF, things could be quite different
if we take into account of the magnetic effect on $Y_{p}$.  This paper will be very useful in the
future study on direct Urca reactions in the magnetar circumstances.

Modified Urca reactions are usually referred to as the standard
cooling process (see Ref.~\refcite{Yakovlev01}), in which the energy loss caused by
resulting neutrinos plays an important role in the thermal
evolution of a NS.  Observations show that the surface thermal temperatures of most
of high-magnetic field pulsars (see Ref.~\refcite{Zhu14}) are higher than those of
ordinary radio pulsars with same characteristic ages. It is found that, the
surface temperatures of magnetars (see Ref.~\refcite{Kaminker06},~\refcite{Mereghetti08},~\refcite{Kaminker14},~\refcite{Popov13},~\refcite{Gao16}~\refcite{Liu16})
are also significantly higher than those of ordinary radio pulsars (see Ref.~\refcite{Popov13},~\refcite{Popov15}).
These observational findings not only challenge the standard cooling theory but also
strongly hint that the magnetic field evolution and the thermal evolution of a magnetar are
closely related. The results of this paper will facilitate the theoretical
research on magneto-thermal evolution of magnetars, facilitate NS other properties research, such as superfluid,hyperons,  asymmetry of nuclear matter and rotational evolution (see Ref.~\refcite{Long12},\refcite{Li12},\refcite{Qi15},\refcite{Li16}).

It is expected that in the near future, our results will
contribute to improving the standard model of neutron star cooling (see Ref.~\refcite{Yakovlev01}), and
will be tested and developed by comparing the improved neutron star
cooling model with magnetar spectrum observations
\section*{Acknowledgments}
We thank anonymous referee for carefully reading
the manuscript and providing valuable comments that improved this paper substantially.
We also thank Prof. Shuang-Nan Zhang for useable discussions, Dr. Raid Yuan for smoothing the language.
This work was supported by Xinjiang Radio Astrophysics Laboratory through grant No. 2015KL012. This work was also supported in part by Xinjiang Natural Science Foundation No.2013211A053, Chinese National
Science Foundation through grants No.11173041,11173042,11003034, 11273051,
11373006 and 11133001, National Basic Research Program of China
grants 973 Programs 2012CB821801 and 2015CB857100, the Strategic Priority Research
Program ``The Emergence of Cosmological Structures'' of Chinese
Academy of Sciences through No.XDB09000000, and the West Light
Foundation of Chinese Academy of Sciences Nos.2172201302, XBB201422, and by a research fund from the Qinglan project of Jiangsu Province.


\section*{References}
\begin{thebibliography}{0}
\bibitem{Thompson96} C. Thompson and R. C. Duncan, {\it Astrophys. J.} {\bf 473} 322 (1996).
\bibitem{Colpi00} M. Colpi, U. Geppert and D. Page, {\it Astrophys. J. Lett.} {\bf 529} 29 (2000).
\bibitem{Woods04}, P. M. Woods, C. Thompson, {\it Compact Stellar X-ray Sources}, (Edited by  W. H. G Lewin and M. van der Klis, {\bf 547}  547 (2004).
\bibitem{Gao13} Z. F. Gao, N. Wang, Q. H. Peng and et al.,{\it Mod. Phys. Lett A.} {\bf 28} 1350138 (2013).
\bibitem{Lam14}Yi Hua. Lam, et al.,{\it EPJ Web of Conferences} {\bf 66} 07011 (2014).
\bibitem{Martinez-Pinedo14}G. Mart\'{i}nez-P\'{i}nedo, Y H. Lam, K. Langanke., et al., {\it Phys. Rev. C.} {\bf 89}  045806 (2014).
\bibitem{Li16}X. H. Li, Z. F. Gao, X. D. Li and et al.,{\it Int. J. Mod. Phys. D.} {\bf 25(1)} 1650002 (2016)
\bibitem{Shapiro83} S. L. Shapiro and S. A. Teukolsky, {\it Black Holes, White Drarfs, and Neutron Stars}, (New York, Wiley-Interscience, 1983)
\bibitem{BPS71} G. Baym, C. Pethick and P. Sutherland, {\it Astrophys. J.} {\bf 170} 299 (1971).
\bibitem{Ruster06}S. B. Ruster, M. Hempel and J. Schaffnet-Bielich, {\it Phys. Rev. C.} {\bf 73} 035804 (2006).
\bibitem{Tsuruta09}S. Tsuruta, J. Sadino, A. akatsuka and  et al., {\it Astrophys. J.} {\bf 691} 621 (2009)
\bibitem{Das12} U. Das and  B. Mukhopadhyay, {\it Phys. Rev. D.} {\bf 86}  042001 (2012).
\bibitem{Dong13}J. Dong, H. Zhang, L. Wang and W. Zuo, {\it Phys. Rev. C.} {\bf 87}  014302  (2013)
\bibitem{Gao15} Z. F. Gao, N.Wang, Y. Xu, et al.,{\it Astron. Nachr.} {\bf 336} No.8/9, 866 (2015)
\bibitem{Ferrer15} E. J. Ferrer, V. de la. Incera, D. M. Paret and et al., {\it Phys. Rev. D.} {\bf 91} 085041 (2015).
\bibitem{Gao11a} Z. F. Gao, N. Wang, D.L. Song and et al., {\it Astrophys. Space Sci.} {\bf 334} 281 (2011).
\bibitem{Gao12a} Z. F. Gao, Q. H. Peng, N. Wang and et al., {\it Astrophys. Space Sci.} {\bf 342} 55 (2012).
\bibitem{Lai91}D. Lai and S. L. Shapiro, {\it Astrophys. J.} {\bf 383} (1991) 745.
\bibitem{Harding06} A. K. Harding and D. Lai, {\it Rep. Prog. Phys.} {\bf 69} (2006) 2631
\bibitem{Landau65} L. D. Landau and E. M.Lifshitz, {\it Quantum Mechanics}, ed. W. H. Freeman, (Pergamon Press, New York, 1965).
\bibitem{Canuto77} V. Canuto and J. Ventura, {\it Fund. Cosmic Phys.} {\bf 2} 203 (1977).
\bibitem{Peng07} Q. H. Peng and H. Tong, {\it Mon. Not. R. Astron. Soc.} {\bf 378} 159 (2007).
\bibitem{Kubo65} R. Kubo, {\it Statistics Mechanics}, (Amsterdam: North-Holland Publishing Co. 1965)
\bibitem{Pauli25} W. Pauli, Z. Physik, {\bf 31},765 (1925).
\bibitem{Pathria03}R. K. Pathria, {\it Statistics Mechanics, 2nd}, (Singapore: Isevier. 2003).
\bibitem{Chakrabarty97} S. Chakrabarty, D. Bandyopadhyay and S. Pal, {\it Phys. Rev. Lett.} {\bf 78} 75 (1997).
\bibitem{Wang14} P. Wang and W. Zuo,  {\it Phys. Rev. C.} {\bf 89} 054319 (2014).
\bibitem{Wang15} P. Wang and W. Zuo,  {\it Chin. Phys. C.} {\bf 39} 014101 (2015).
\bibitem{Gao11b} Z. F. Gao, N. Wang, D.L. Song and et al., {\it Astrophys. Space Sci.} {\bf 334} 281 (2011).
\bibitem{Gao12b}Z. F. Gao, Q. H. Peng, N. Wang and et al., {\it Chin. Phys. B.} {\bf 21} 057109 (2012).
\bibitem{Yakovlev01} D. G. Yakovlev, A. D. Kaminker, O. Y. Gnedin and et al. {\it Phys. Rep.} {\bf 354} 1 (2001).
\bibitem{Boguta77} J. Boguta and A. R. Bodmer {\it Nuclear. Physics. A.} {\bf 292} 413 (1977).
\bibitem{Boguta81} J. Boguta. {\it Phys. Lett.} {\bf 106} 255 (1981).
\bibitem{Boguta83} J. Boguta and H. Stocker {\it  Physics. Lett.} {\bf 120} 289 (1983).
\bibitem{Takahashi08}K. Takahashi, {\it Journal of Physics G-Nuclear and Particle Physics}, {\bf 35(3)} 035002 (2008).
\bibitem{Henley07} E. M. Henley, M. B. Johnson and L. S. Kisslinger, {\it  Phys. Rev. D.} {\bf 76} 125007 (2007).
\bibitem{Baiko99} D. A. Baiko and D. G. Yakovlev, {\it Astron. Astrophys.}{\bf 342} 192 (1999).
\bibitem{Zhu14} H. Zhu, W. W. Tian and P. Zuo, {\it Astrophys. J.} {\bf 793} 95 (2014).
\bibitem{Kaminker06} A. D. Kaminker, D. G. Yakovlev, A. Y. Potekhin and et al., {\it Mon. Not. R. Astron. Soc.} {\bf 371} 477 (2006).
\bibitem{Mereghetti08} S. Mereghetti, {\it Astron. Astrophys. Rev. } {\bf 15} 225 (2008).
\bibitem{Kaminker14} A. D. Kaminker, A. A. Kaurov, A. Y. Potekhin and D. G. Yakovlev, {\it Mon. Not. R. Astron. Soc.} {\bf 442} 3484 (2014).
\bibitem{Gao16} Z. F. Gao, X. D. Li, N. Wang, et al., {\it Mon. Not. R. Astron. Soc.} {\bf 456} 55 (2016).
\bibitem{Liu16} Jing-Jing Liu, {\it Research in Astronomy and Astrophysics.} (2016) arXiv:1602.05501
\bibitem{Popov13} S. B. Popov, {\it Publications of the Astronomical Society of Australia} {\bf 30} 045 (2013).
\bibitem{Popov15} S. B. Popov, A. A. Kaurov and A. D. Kaminker, {\it Publications of the Astronomical Society of Australia} {\bf 32} 018 (2015)
\bibitem{Long12} W.H. Long, et al., {\it Phys. ReV. C.}, {\bf 85} 025806 (2012)
\bibitem{Li12} Z. H. Li and W. Zuo, {\it Phys. ReV. C.}, {\bf 85} 037001 (2012)
\bibitem{Qi15} B. Qi, et al., {\it Chin. Phys. Lett.} {\bf 32} 112101 (2015)
\bibitem{Li16} A. Li, et al., {\it Phys. ReV. C.}, {\bf 93} 015803 (2016)
\end{thebibliography}
\end{document}